\documentclass[prb,twocolumn]{revtex4}

\makeatletter
\def\@dotsep{4.5}
\makeatother

\usepackage{epsfig}

\begin{document}

\title{Boltzmann bias grand canonical Monte Carlo}

\author{G. Garberoglio}
\email{garberog@science.unitn.it}
\affiliation{Dipartimento di Fisica dell'Universit\`a di Trento and CNISM, via
  Sommarive 14, I-38100 Trento (TN), Italy}

\begin{abstract}
We derive an efficient method for the insertion of structured particles in
grand canonical Monte Carlo simulations of adsorption in very confining
geometries. We extend this method to path integral simulations and use it to
calculate the isotherm of adsorption of hydrogen isotopes in narrow carbon
nanotubes (2D confinement) and slit pores (1D confinement) at the temperatures
of $20$~K and $77$~K, discussing its efficiency by comparison to the standard
path integral grand canonical Monte Carlo algorithm.

We use this algorithm to perform multicomponent simulations in order to
calculate the hydrogen isotope selectivity for adsorption in narrow carbon
nanotubes and slit pores at finite pressures.

The algorithm described here can be applied to the study of adsorption of real
oligomers and polymers in narrow pores and channels.
\end{abstract}

\date{\today}

\maketitle

\section{Introduction}

The grand canonical (GC) Monte Carlo (MC) method is an efficient way to
calculate thermodynamic properties of adsorbed fluids via computer
simulations.~\cite{at,binder} It has recently been extended to take into
account quantum diffraction effects using the path integral (PI) formulation
of quantum mechanics if the adsorption of light particles, such as helium
atoms or hydrogen molecules, is investigated at low
temperatures.~\cite{pigcmc,slit-pores}

Use of the method for the study of hydrogen adsorption in very confining
geometries has led to the discovery of a very strong isotope
effect.~\cite{KatorskiW64,Moiseyev75} The adsorption of hydrogen isotopes in
carbon nanotubes has been thoroughly investigated as a function of 
temperature and nanotube radius, and it has been found that these
nanostructured materials might be used, in principle, to obtain a very
efficient isotope separation, a phenomenon referred to as ``quantum
sieving'' and studied theoretically for a long
time.~\cite{krylov,wang,pigcmc,challa_zerop,challa,hathorn,trasca,goldfield2006,Garberoglio2006,slit-pores} 
Recently the first experimental confirmations have been reported.~\cite{sieving-expt,sieving-expt2}

From a simulation point of view the PIMC method maps the statistical
mechanics of $N$ quantum particles into the statistical mechanics of $N$ ring
polymers, each formed by $P$ particles. The mapping is exact in the limit $P
\rightarrow \infty$, even though, according to the temperature investigated,
convergence of the results can be obtained when $P$ is of the order of
10-100.

The standard path integral grand canonical (PIGCMC) simulation procedure was
developed by Johnson and collaborators~\cite{pigcmc} who used ring polymer
configurations drawn from an ideal gas distribution as insertion candidates.
While this method is effective for studying adsorption in large pores, it is
very inefficient when applied to very narrow geometries. The reason of this
inefficiency stems from the fact that in a strongly confined system at low
temperatures the average kinetic energy of the adsorbed particles might be
much higher than the average thermal kinetic energy of a free particle.  In
the path integral formulation this quantity is related to the size of the ring
polymers: polymers are smaller - on average - when they represent particles
with higher kinetic energy.  As a consequence, if one tries to insert ring
polymers drawn from a ideal gas distribution at the desired temperature in a
strongly confining geometry, the likelihood of a rejection is very high,
because the ``average'' ideal gas configuration just does not fit in the
available space within the adsorbent.

This phenomenon has been recognized in the early PIGCMC simulations, since very
low acceptance ratios were observed when trying to insert ideal gas ring
polymers in very narrow carbon nanotubes.~\cite{challa} The same phenomenon
has also been observed by independent studies in the case of adsorption in
carbon slit pores and and has been called ``quantum polymer shrinking'' by
Kowalczyk {\it et al.}~\cite{slit-pores} and it is a consequence of the
well known fact that more confining geometries enhance the zero point energy
of adsorbed particles. This effect is even more pronounced when the rotational
degrees of freedom of the hydrogen molecule are taken into account, leading to
a dramatic enhancement of the sieving
properties.~\cite{Garberoglio2006,goldfield2006,Garberoglio2007}

We propose a new algorithm for insertion moves in classical and quantum (path
integral) grand canonical Monte Carlo simulations, which is particularly
suitable in the case of strong confinement. We show that the acceptance ratio
of insertion moves can be up to two orders of magnitude higher than the one
obtained using ideal gas ring polymers as trial configurations. As a
consequence the equilibration and production runs of a PIGCMC simulation can
be made correspondingly shorter, considerably reducing the time needed for the
calculations.

This method, which we call Boltzmann bias grand canonical Monte Carlo, uses as
trial configurations ring polymers obtained by an independent path integral
Monte Carlo simulation of particles which are adsorbed within the same
adsorbent, but do not interact between themselves, so that the ``shape'' of
the ring polymers to be inserted is already the correct one and consequently
the insertion candidates are more likely to fit within the confining geometry.

In the following we derive the expressions for the trial and acceptance
probability, both in the classical and in the quantum case. Although the
Boltzmann bias method is not likely to produce any improvement in classical
simulations of structureless particles with respect to other existing
algorithms, we present the derivation in detail for the classical case. This
will be done to fix the notation before passing on to the quantum
case. We also present this algorithm in the case of adsorption of oligomers
and polymers within confining geometries.

We will apply the Boltzmann bias method to the calculation of adsorption
isotherms and finite pressure selectivities for mixtures of hydrogen isotopes
in narrow carbon nanotubes and slit pores, and compare our results with those
obtained using the standard PIGCMC method.

The analysis of the efficiency of the Boltzmann bias method will show that one
can obtain acceptance ratios up to four orders of magnitude higher than the
ones obtained using the standard insertion algorithm, dramatically reducing
the amount of time needed to perform the computer simulations.

\section{Boltzmann bias method}

\subsection{Classical simulations}

The GCMC method is based on the generation of a Markov chain having as
equilibrium probability density the quantity
\begin{equation}
P^{(\rm eq)}(x_1,\ldots,x_N) = \frac{1}{\Xi} \frac{e^{\beta \mu N}}{N!
  \Lambda^{3N}} e^{-\beta U(x_1,\ldots,x_N)} ~ d^3x_1 \ldots d^3x_N
\end{equation}
where $N$ is the variable number of particles, and $V$, $T$ and $\mu$ are the
system's volume, the temperature and the chemical potential, respectively,
which are all kept fixed.  The quantity $\Lambda = h / \sqrt{2 \pi m k_B T}$
is the thermal de Broglie wavelength of the particles of mass $m$ and $\Xi =
\sum_N e^{\beta \mu N} Q_N(V,T)$ is the grand canonical partition function.
$Q_N(V,T)$ is the canonical partition function,$\beta = 1/k_B T$ and $k_B$ is
the Boltzmann constant.  Finally, $U(x_1,\ldots,x_N)$ is the interaction
potential energy between the $N$ particles and/or between the particles and an
external potential (which can model, for example, an adsorbent).

The Markov chain is built by moves that either keep the number of particles
constant (and for which the standard canonical Monte Carlo algorithms can be
used) and moves that change the number of particles in the system. It is
usually sufficient to consider moves in which the number of particles is
changed by unity.

The expression for the transition probability $W_{i \rightarrow j}$ of the
Markov chain between states $i$ and $j$ having $N$ and $N+1$ particles
respectively can be obtained by applying, as usual, the detailed balance
condition, i.e. 
\begin{equation}
\frac{W_{N \rightarrow N+1}}{W_{N+1 \rightarrow N}} = 
\frac{P^{(\rm eq)}_{N+1}}{P^{(\rm eq)}_N} = 
\frac{e^{\beta \mu}}{(N+1) \Lambda^3} e^{- \beta \Delta U} ~ d^3x_{N+1}
\label{eq:detailed-balance}
\end{equation}
where $\Delta U = U_{N+1} - U_N$ is the difference between the potential
energy of the two states $i$ and $j$. The transition probability $W$ is
written as the product of a trial probability $T$ and an acceptance
probability $A$. In the standard grand canonical Monte Carlo algorithm the
trial probability for an insertion is such that a test particle is inserted
uniformly within the volume and the the probability of a removal is just one,
hence
\begin{equation}
\begin{array}{rcl}   
T_{N \rightarrow N+1} &=& \displaystyle\frac{ d^3x_{N+1} }{V} \\
T_{N+1 \rightarrow N} &=& 1
\end{array}
\label{eq:standard-trial}
\end{equation}
so that the ratio between the acceptance probabilities corresponding to the
insertion into and removal from the system of a particle is
\begin{equation}
\frac{A_{N \rightarrow N+1}}{A_{N+1 \rightarrow N}} =
\frac{V}{(N+1) \Lambda^3} e^{\beta \mu} e^{- \Delta U},
\end{equation}
a condition that can be satisfied, e.g., by the standard Metropolis choice
\begin{equation}
\begin{array}{rcl}
A_{N \rightarrow N+1} &=& \min \left[ 1,
\displaystyle\frac{V}{(N+1) \Lambda^3}~e^{\beta \mu} e^{- \Delta U} \right] \\
A_{N+1 \rightarrow N} &=& \min \left[ 1,
\displaystyle\frac{(N+1) \Lambda^3}{V} ~ e^{-\beta \mu} e^{\Delta U} \right]
\end{array}
\label{eq:Metropolis}
\end{equation}

The Boltzmann bias methods stems now from the fact that in many cases the
interaction potential energy can be written as the sum of the potential energy
of interaction with the fluid particles between themselves ($U_{\rm FF}$) and
the interaction potential energy with a substrate ($U_{\rm SF}$), the
latter being additive in the number of particles, i.e.
\begin{widetext}
\begin{equation}
U({x_1,\ldots,x_N}) = 
U_{\rm FF}(x_1,\ldots,x_N) + U_{\rm SF}(x_1,\ldots,x_N) = 
U_{\rm FF}(x_1,\ldots,x_N) + \sum_{k=1}^N u_{\rm SF}(x_k)
\end{equation}
\end{widetext}
so that one can devise an insertion move such that the probability of
insertion within the volume $V$ is not uniform, but proportional to the
Boltzmann factor of the solid-fluid interaction potential of the particle to
be inserted, $\exp\left[ -\beta u_{\rm SF}(x_{N+1})\right]$.
One can then write the trial probability for an insertion as
\begin{widetext}
\begin{equation}
T_{N \rightarrow N+1} = 
\frac{\exp\left[{-\beta u_{\rm SF}(x_{N+1})}\right] ~ d^3x_{N+1}}
{\displaystyle\int \exp\left[{-\beta u_{\rm SF}(x_{N+1})}\right] ~ d^3x_{N+1}}
\equiv 
\frac{\exp\left[{-\beta u_{\rm SF}(x_{N+1})}\right] ~ d^3x_{N+1}}{V~ e^{-\beta
    \bar\mu}}
\label{eq:bb-cl-trial}
\end{equation}  
\end{widetext}
where the last equality defines the quantity $\bar\mu$. With this choice for
the trial probability of an insertion (leaving $T_{N+1 \rightarrow N}
= 1$), one gets for the acceptance probability the ratio
\begin{equation}
\frac{A_{N \rightarrow N+1}}{A_{N+1 \rightarrow N}} =
\frac{V}{(N+1) \Lambda^3} e^{\beta (\mu - \bar\mu)} e^{- \Delta U_{\rm FF}(x_1,\ldots,x_{N+1})}
\label{eq:bb-classical}
\end{equation}
that depends only on the difference of the fluid-fluid potential energy
between the original configuration and the trial one.

One way to insert particles with the trial probability given by
Eq.~(\ref{eq:bb-cl-trial}) is to perform, in parallel with the grand canonical
simulation, a standard canonical Monte Carlo simulation of some particles
within a similar box and at the same temperature. In this second simulation
the particles interact only with the substrate and not among
themselves, so that at equilibrium they sample the distribution given by
Eq.~(\ref{eq:bb-cl-trial}). One can then get the coordinates for the GCMC
trial insertions by drawing particles' positions from the configurations
generated by this second simulation.

In the case of classical adsorption of structureless particles this method is
not particularly more efficient than other standard methods based on the
insertion in the subvolume of the simulation cell where the solid-fluid
potential energy is less than a given threshold. This free volume $V_{\rm
  free}$ can be easily pre-calculated and used in the standard grand canonical
algorithm for insertions, Eq.~(\ref{eq:standard-trial}), instead of the whole
cell volume $V$.

In the case of classical adsorption of molecules, the detailed balance
condition (\ref{eq:detailed-balance}) can be written as
\begin{widetext}
\begin{equation}
\frac{W_{N \rightarrow N+1}}{W_{N+1 \rightarrow N}} = 
\frac{e^{\beta \mu}}{(N+1) \Lambda^3} e^{- \beta \Delta U_{\rm FF}} 
e^{-\beta U_{\rm SF}(x^{\rm CM}_{N+1}, \xi)} 
e^{-\beta U_{\rm int}(\xi)}
~ d^3x^{\rm CM}_{N+1} d \xi
\label{eq:db-cl-mol}
\end{equation}  
\end{widetext}
where $x^{\rm CM}_{N+1}$ denotes the center of mass position and $\xi$ are the
variables describing the intra-molecular degrees of freedom (such as Euler
angles in the case of small rigid molecules or torsion angles for small
oligomers). The quantity $U_{\rm int}(\xi)$ is just the internal energy
corresponding to the configuration given by the coordinates $\xi$.
One can see from Eq.~(\ref{eq:db-cl-mol}) that the transition probability
depends on the internal state of the molecule $\xi$ as well as the position of
the center of mass for the trial insertion.
In the limit of low density the fluid-fluid contribution to the acceptance 
probability can be neglected and this quantity is mainly determined by the
solid-fluid part of the interaction.
In the case of adsorption in narrow geometries, many of the trial
configurations $\xi$ for a given value of the center of mass position $x^{\rm
  CM}_{N+1}$ are likely to be rejected, since they would result in strong
overlaps of parts of the molecule with the adsorbent.
In this case a much greater acceptance ratio could be obtained if the
configurations $\xi$ were chosen among the ones most likely to fit at a given
position $x^{\rm CM}_{N+1}$. 
The Boltzmann bias method proposed here is based on choosing the
the trial configurations with a probability proportional to the Boltzmann
factor of the molecule to be inserted at the position $x^{\rm CM}_{N+1}$, i.e.
\begin{equation}
T_{N \rightarrow N+1} \propto e^{-\beta U_{\rm SF}(x^{\rm CM}_{N+1}, \xi)}
e^{-\beta U_{\rm int}(\xi)} ~ d^3x^{\rm CM}_{N+1} d \xi.
\label{eq:bb-cl-mol-trial}
\end{equation}
By writing the normalization factor as
\begin{equation}
V e^{\beta \bar\mu} = \int
e^{-\beta U_{\rm SF}(x^{\rm CM}_{N+1}, \xi)} e^{-\beta U_{\rm int}(\xi)}
~ d^3x^{\rm CM}_{N+1} d \xi
\label{eq:cl-mol-mubar}
\end{equation}
one obtains for the acceptance probability an expression analogous to the one
reported in Eq.~(\ref{eq:bb-classical}).
The trial configurations can be easily generated by running an $NVT$
simulation of molecules within the adsorbent, with the fluid-fluid interaction
switched off.

Since the solid-fluid interaction has already been taken into consideration
by the choice of the trial probability, the acceptance probability now depends
on the values of the fluid-fluid interaction between the molecule to be
inserted and the ones already present within the simulation box, as well as on
the value of the renormalized chemical potential $\mu - \bar\mu$. One can then
expect a very high acceptance ratio for all the thermodynamic state points
where the density of the adsorbed molecules is not liquid-like.

\subsection{Path integral simulations}

The path integral grand canonical method is based on the Trotter factorization
of the many-body Hamiltonian. The quantum partition function of $N$
interacting particles can be written as
\begin{equation}
Q_N(V,T) = \frac{1}{N!} 
\sum_\pi \langle \exp\left[ -\beta (\hat T + \hat V ) \right] P_\pi \rangle
\end{equation}
where $\hat T$ and $\hat V$ are the kinetic and potential energy operators,
respectively, and the expectation value is taken over a complete set of $N$
particle states. The sum is over all the permutations $\pi$ of the particles,
represented in the Hilbert space by the operator $P_\pi$ which also takes into
account the bosonic or fermionic nature of the particles. In the
following we will not consider effects due to quantum statistics and hence we
will approximate the sum with its leading term, corresponding to the identity
permutation. This is a good approximation as long as the
thermal de Broglie wavelength $\Lambda$ does not exceed the hard core radius
of the particles under consideration.~\cite{rpf-statmech}
By using the Trotter formula
\begin{equation}
e^{\hat A + \hat B} = 
\lim_{P \rightarrow \infty} \left( e^{\hat A / P} e^{\hat B/P} \right)^P
\end{equation}
with a finite but large enough Trotter index $P$, the quantum partition
function of $N$ particles can be mapped into the classical partition function
of $NP$ particles, such that each original particle corresponds to a ring
polymer of $P$ ``beads''.~\cite{ceperley} The beads within each polymer
interact with their two nearest neighbors via an harmonic interaction, and we
will call this part of the interaction ``internal''.  Denoting by $X_i =
\{x^{(1)}_i, \ldots, x^{(P)}_i \}$ the $3P$ coordinates of the beads of the
ring polymer corresponding to the $i$-th particle, and applying the same
reasoning as described before in the classical case, one obtains for the path
integral case
\begin{widetext}
\begin{equation}
\frac{W_{N \rightarrow N+1}}{W_{N+1 \rightarrow N}} = 
\frac{1}{N+1} \left( \frac{P^{3P/2}}{\Lambda^{3P}} \right) 
e^{\beta \mu}
e^{- \beta \Delta \overline{U_{\rm FF}}(X_1,\ldots,X_{N+1})} 
e^{- \beta \overline{U_{\rm SF}}(X_{N+1})} 
e^{- \beta U_{\rm int}(X_{N+1})} 
~ d^3X_{N+1}
\end{equation}  
\end{widetext}
where we have defined
\begin{eqnarray}
\overline{U_{\rm FF}}(X_1,\ldots,X_{N}) &=& 
\frac{1}{P}  \sum_{p=1}^P \sum_{i<j=1}^N
u_{\rm FF}(|x^{(p)}_j - x^{(p)}_i|)
\\
\overline{U_{\rm SF}}(X_{N+1}) &=& 
\frac{1}{P} \sum_{p=1}^P u_{\rm SF}(x^{(p)}_{N+1}) \\
U_{\rm int}(X_{N+1}) &=& k_B T ~ \frac{K}{2} \sum_{p=1}^P 
\left| x^{(p)}_{N+1} - x^{(p+1)}_{N+1} \right|^2,
\end{eqnarray}
$K = 2 \pi P / \Lambda^2$, $\Delta \overline{U_{FF}}(X_1,\ldots,X_{N+1})
= \overline{U_{\rm FF}}(X_1,\ldots,X_{N+1}) - \overline{U_{\rm
    FF}}(X_1,\ldots,X_{N}) $ and we have used the convention
$x^{(P+1)}_{N+1} = x^{(1)}_{N+1}$. In the path integral GCMC case the ratio
between the transition probabilities connecting states with different number
of particles depends also on the internal state of the added polymer and it is
customary to draw internal states from an ideal gas path integral
simulation.~\cite{pigcmc}

This way of proceeding is highly inefficient in the case of strong
confinement, where one expects that the distribution of shapes for the
adsorbed ring polymers, having to account for a high kinetic energy due to
Heisenberg indetermination, will be considerably different from the
equilibrium distribution of an ideal gas. In particular it turns out that
polymers corresponding to states of higher kinetic energy tend to have a
smaller size.~\cite{slit-pores} As a consequence the probability of
rejecting an insertion move is particularly high, because almost all of the
ideal gas ring polymer configurations do not easily fit into the available
space within the adsorbent.

In this case the Boltzmann bias method is likely to improve the
situation. Instead of drawing a ring polymer configuration from an ideal gas
distribution, the trial insertion probability is chosen to be proportional to
the probability of finding an isolated ring polymer within the adsorbent, i.e.
\begin{equation}
T_{N \rightarrow N+1} \propto 
\left( \frac{P^{3P/2}}{\Lambda^{3P}} \right) 
e^{- \beta \overline{U_{\rm SF}}(X_{N+1})} 
e^{- \beta U_{\rm int}(X_{N+1})} 
~ d^3X_{N+1} 
\end{equation}
By using the change of variables
\begin{equation}
  \begin{array}{ccc}    
  r &=& x^{(1)}_{N+1} \\
  \Delta r_1 &=& x^{(2)}_{N+1} - x^{(1)}_{N+1}  \\
  &\cdots& \\
  \Delta r_{P-1} &=& x^{(P)}_{N+1} - x^{(P-1)}_{N+1}
  \end{array}
\label{eq:change_var}
\end{equation}
the normalization factor can be written as the integral
\begin{widetext}
\begin{equation}
\frac{1}{\Lambda^3}\int e^{- \beta \overline{U_{\rm SF}}(r,\Delta r_1, \ldots, \Delta r_{P-1})}
F_{\rm ring}(\Delta r_1, \ldots, \Delta r_{P-1}) ~\prod_{p=1}^{P-1} 
  \Delta r_p ~ d^3r =
V e^{-\beta \bar\mu}
\label{eq:qu-mubar}
\end{equation}
\end{widetext}
where $F_{\rm ring}(\Delta r_1, \ldots, \Delta r_{P-1})$ is the probability of
having a configuration of an ideal gas ring polymer with separations $\Delta
r_p$ between the neighboring beads, and it is derived in
Appendix~\ref{sec:appendix}.  The last equality defines the quantity $\bar\mu$
analogously to the classical definition given in Eq.~(\ref{eq:bb-cl-trial}).

With these definitions the ratio of the acceptance probabilities for moves
which change the number of particles in the path integral case is
\begin{equation}
\frac{A_{N \rightarrow N+1}}{A_{N+1 \rightarrow N}} =
\frac{V}{(N+1) \Lambda^3} e^{\beta (\mu - \bar\mu)} e^{- \Delta \overline{U_{\rm
    FF}}(X_1,\ldots,X_{N+1})}
\label{eq:bb-quantum}
\end{equation}
completely analogous to the classical case of Eq.~(\ref{eq:bb-classical}).
By ``undoing'' the Trotter factorization in Eq.~(\ref{eq:qu-mubar}) one can
show that
\begin{equation}
V e^{-\beta \bar\mu} = 
\Lambda^3 \left\langle 
\exp \left[ -\beta \left( \frac{\hat p^2}{2m} + u_{\rm SF}(\hat x) 
\right) \right] \right\rangle
= \Lambda^3 \sum_i e^{-\beta E_i}
\label{eq:mubar}
\end{equation}
where $E_i$ are the quantum energy levels of a single particle interacting
with the adsorbent.

The Boltzmann bias PIGCMC method is particularly efficient
in the case where adsorption is simulated in materials having an (approximate)
continuous translational symmetry as it happens, e.g., in carbonaceous
materials such as carbon nanotubes and slit pores, where it has been shown
that the corrugation of the potential energy can be neglected without
sacrificing accuracy.~\cite{challa_zerop} In this case the ring polymer
configurations for trial insertions can be further translated by a random
value along the direction(s) of translational symmetry.
This prescription avoids the need to wait for the polymers in the
zero-pressure simulation box to diffuse away from positions already used for
trial insertions.


In the case of simulations in geometries that do not possess continuous
translational symmetry, as it might happen if the effect of nanotube
corrugation on adsorption had to be addressed or in the case of adsorption in
amorphous materials, one should perform long stages of $NVT$ MC in the
zero-pressure box until a new configuration, uncorrelated with the previous
one, is generated. This is necessary in order for the new polymers used for
trial insertions not to end up on top of polymers already present in the 
simulation box, resulting in a certain rejection.

In the general case another method for generating trial polymers, which seems
particularly suited for path integral simulations, could be used. One starts
from configurations where $N$ independent classical point particles are
adsorbed at the target temperature $T$ (which can be generated very
efficiently, even in a parallel environment, either by GCMC or molecular
dynamics), and use them to build $N$ ring polymers where all the beads are at
the position of the corresponding classical particles.  These polymers can be
efficiently equilibrated using hybrid MC~\cite{hmc,pigcmc} and then used as
insertion candidates.  This approach might be more efficient than the direct
one outlined above, and its performance will be analyzed in future studies.

There is another issue to be addressed in the case of simulation of adsorption
in more general geometries, i.e. the calculation of the parameter $\bar \mu$.
In a general geometry the calculation of the energy levels of a single
particle, needed to compute $\bar\mu$ according to Eq.~(\ref{eq:mubar}), can
be very computationally demanding and in this case it could be worth it using
directly Eq.~(\ref{eq:qu-mubar}). In fact the left hand side of
Eq.~(\ref{eq:qu-mubar}) consists of a three dimensional integration over the
variable $r$ of the $e^{- \beta \overline{U_{\rm SF}}(r,\Delta r_1, \ldots,
  \Delta r_{P-1})}$ averaged over ring polymer configurations drawn from an
ideal gas distribution, which can be generated rather
efficiently.~\cite{fosdick-jordan66}


\section{Adsorption in narrow carbon nanotubes}

\subsection{Pure fluid isotherms}

In order to validate the Boltzmann bias method we compare the results with
standard path integral grand canonical Monte Carlo simulations of
adsorption in carbon nanotubes, where insertion moves are performed by taking
ring polymer configurations from an ideal gas simulation performed in parallel
with the main computation.
 
We model the carbon nanotubes as smooth tubes of given geometrical radius
$R$. The solid-fluid interaction depends only on the distance of the particle
from the tube axis, and we used the expression for the average potential
reported in Refs.~\onlinecite{matt,hypgeo1,hypgeo2}. We
chose two values of the geometrical radius: $R = 3.6$~\AA\ that corresponds
to the (2,8) tube and $R = 3.1$~\AA\ that corresponds to the
(3,6) tube.

Earlier path integral simulations of hydrogen adsorption in these tubes showed
that at the temperature of $T = 20$~K, quantum effects are very important to
describe the actual behavior of hydrogen adsorbed in the (3,6), and moderately
important in the case of the (2,8) tube.~\cite{challa_zerop,challa}
At a higher temperature of $T = 77$~K, quantum effects on adsorption in the
(2,8) tube are expected to be smaller.

In our simulations the fluid-fluid interaction between hydrogen atoms is
assumed to be of the Lennard-Jones type, with the Buch
parameters~\cite{Buch94} $\varepsilon_{\rm HH}/k_B = 34.2$~K and $\sigma_{\rm
  HH} = 2.96$~\AA, and the solid-fluid hydrogen-carbon interaction is
calculated using the Lorentz-Berthelot mixing rules~\cite{at} together with
the Steele parameters~\cite{Steele78} for the carbon atom, $\varepsilon_{\rm
  CC}/k_B = 28.0$~K and $\sigma_{\rm CC} = 3.4$~\AA.  The quantity $\bar\mu$
appearing in Eq.~(\ref{eq:bb-quantum}) has been calculated using
Eq.~(\ref{eq:mubar}). The energy levels of a single hydrogen particle
interacting with a carbon nanotube have been obtained by direct
diagonalization of the quantum Hamiltonian, using as a basis set the
eigenfunctions of a free particle confined within a rigid cylinder of radius
$R$. In the case of the (3,6) nanotube, the first 183 energy states of the
free particle were sufficient to reach convergence, whereas in the case of the
(2,8) tube we obtained convergence of the results using 226 states.  The
values of $\bar\mu$ as well as other relevant single particle properties of
hydrogen isotopes confined in carbon nanotubes are reported in
Table~\ref{tab:h2_nt}.  The results of these single particle calculations were
used to fix the value of the Trotter number $P$ as a function of the
temperature: we performed path integral canonical simulations of hydrogen
adsorbed within the tubes switching off the interparticle interaction,
progressively increasing the value of $P$ until we obtained the same average
values for the kinetic and potential energies as those reported in
Table~\ref{tab:h2_nt}. We reached convergence using $P=64$ for H${}_2$ at
$T=20$~K and using $P=16$ for H${}_2$ at $T=77$~K. The same values of the
Trotter index $P$ obtained for H${}_2$ were used for the other isotopes.
\begin{table*}
\begin{tabular}{c|c|c|c|c|c}    
Nanotube type & Adsorbate & Temperature (K) & $\bar\mu/k_B$ (K) & 
Kinetic Energy/$k_B$ (K) & Potential Energy/$k_B$ (K)  \\
\hline
(2,8) & H${}_2$ & 20 & $-1245 $  & 124.0 & $-1392 $\\
(2,8) & T${}_2$ & 20 & $-1314 $  & 69.1  & $-1428 $\\
(2,8) & H${}_2$ & 77 & $-1054 $  & 167.4 & $-1383 $\\
(2,8) & T${}_2$ & 77 & $-1088 $  & 132.5 & $-1404 $\\
\hline
(3,6) & H${}_2$ & 20 & $-281.3$  & 329.9 & $-629.7$\\ 
(3,6) & T${}_2$ & 20 & $-524.3$  & 187.2 & $-751.0$\\ 
(3,6) & H${}_2$ & 77 & $-99.05$  & 358.6 & $-628.6$\\ 
(3,6) & T${}_2$ & 77 & $-281.0$  & 219.5 & $-747.7$
\end{tabular}
\caption{Single particle properties of hydrogen isotopes confined in carbon
  nanotubes, obtained by direct diagonalization of the single particle
  Hamiltonian. The quantity $\bar{\mu}$ is defined in Eq.~(\ref{eq:mubar}).}
\label{tab:h2_nt}
\end{table*}  

\begin{figure}
\epsfig{file=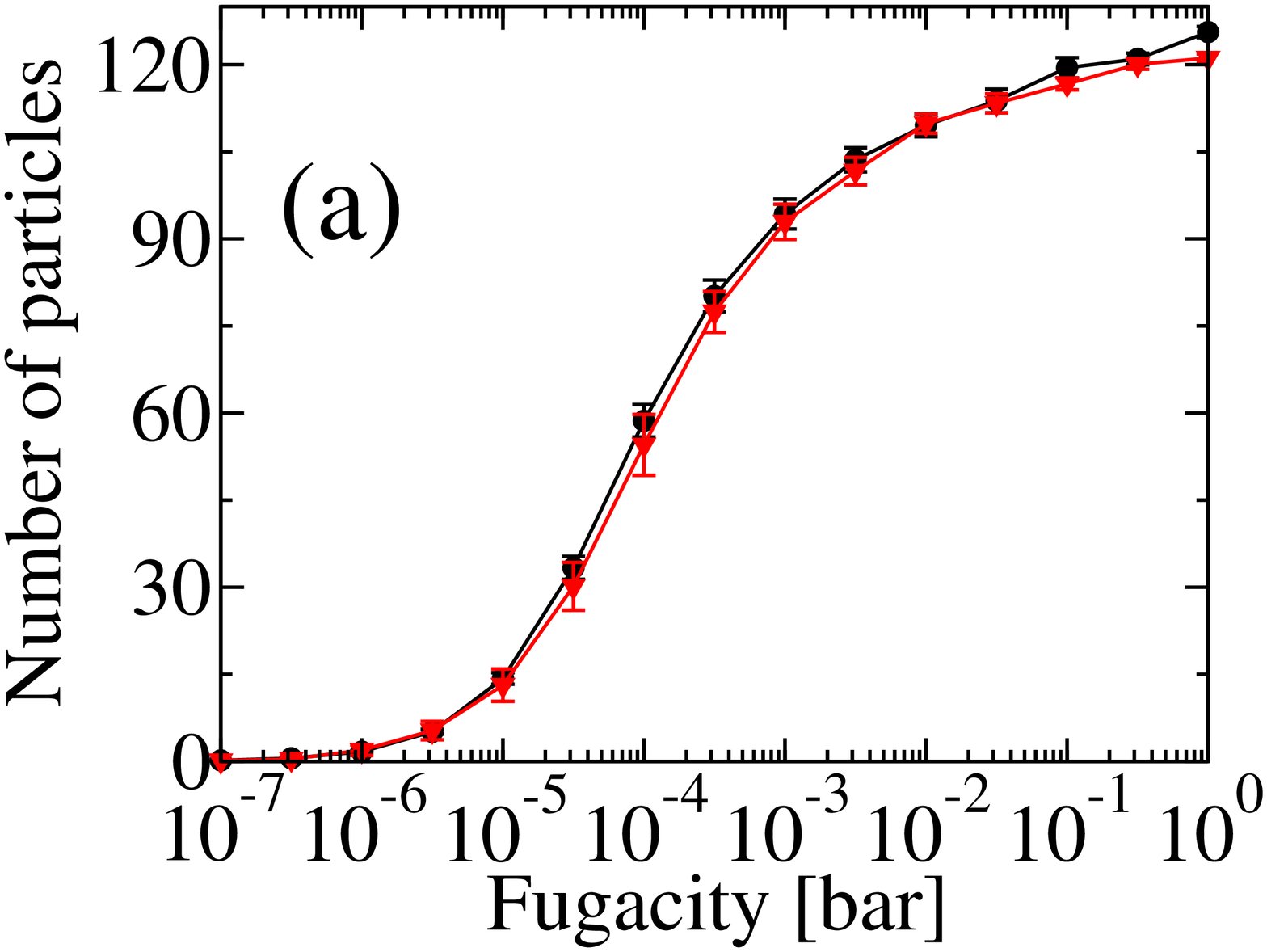,width=0.8\linewidth}
\epsfig{file=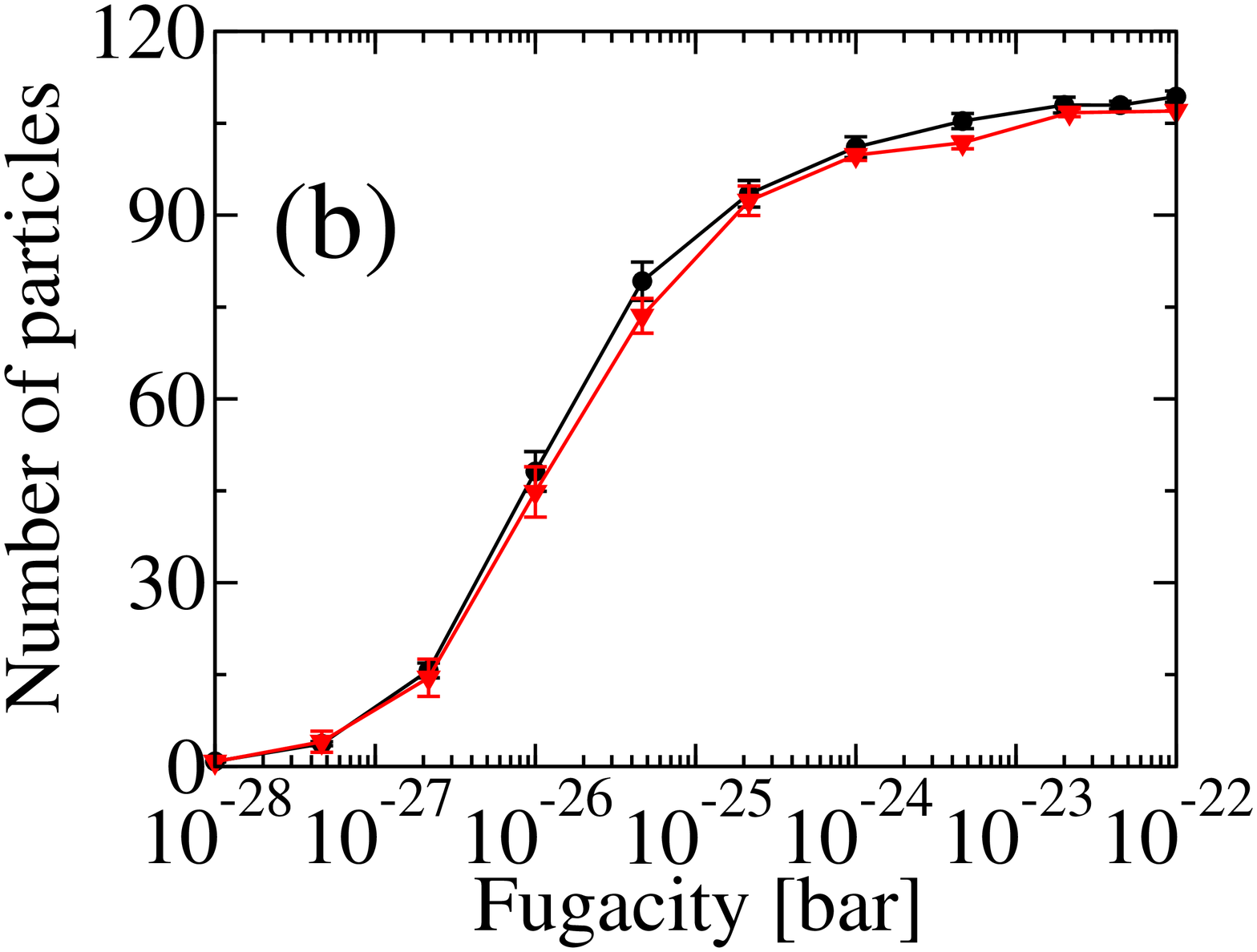,width=0.8\linewidth}
\epsfig{file=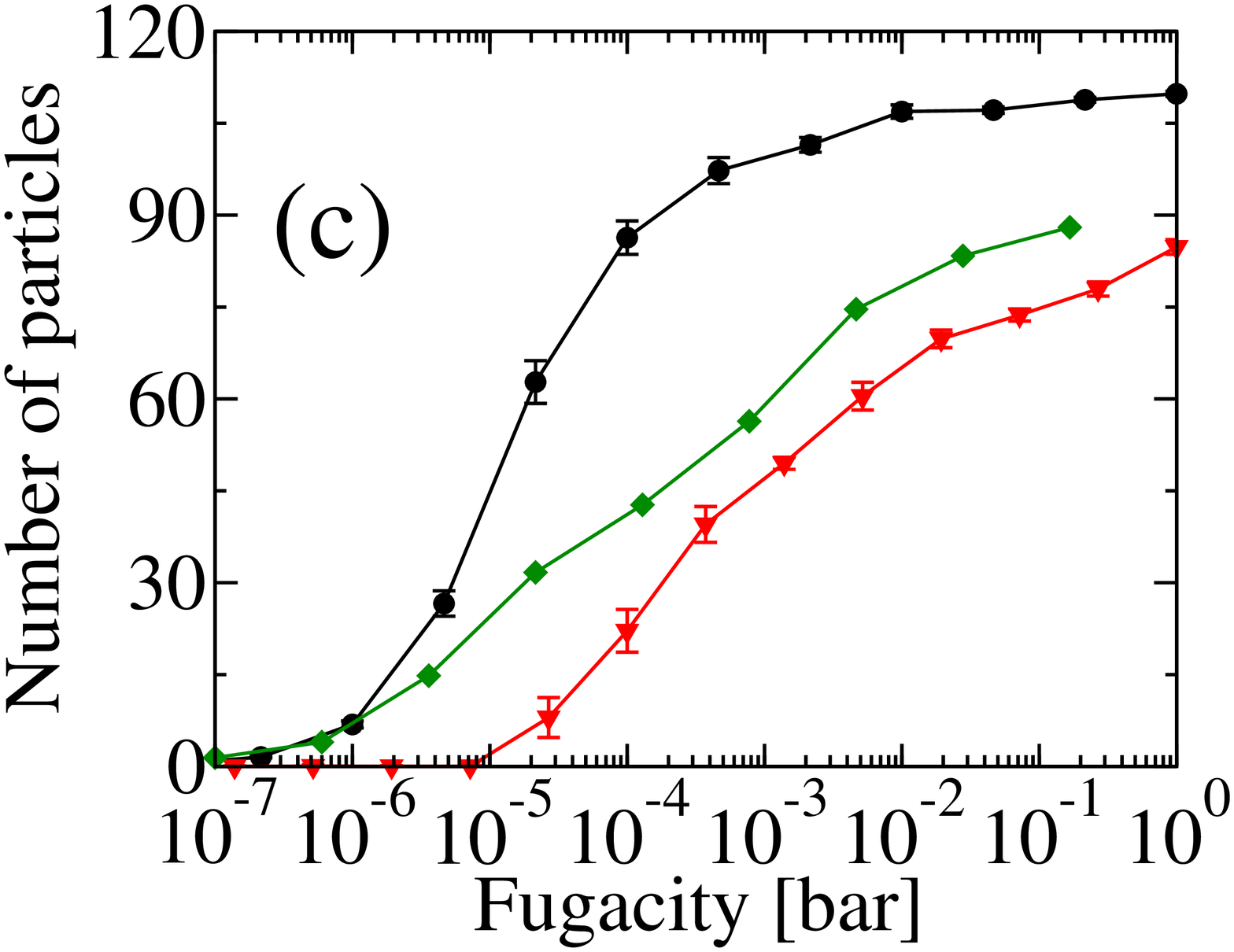,width=0.8\linewidth}
\caption{Isotherms of adsorption for hydrogen in carbon nanotubes. (a) (2,8)
  tube at $T=77$~K, (b) (2,8) tube at $T=20$~K and (c) (3,6) tube at
  $T=20$~K. The circles denote the results using the Boltzmann bias method, 
  whereas the triangles are the results of simulations using ideal gas ring
  polymers as insertion candidates. The diamonds in panel (c) are
  the result of a standard PIGCMC calculation with twice as many MC steps as
  the other one.}
\label{fig:numbers}
\end{figure}

We show in Fig.~\ref{fig:numbers} the adsorption isotherms
obtained using both the Boltzmann bias method and the standard procedure for
path integral grand canonical Monte Carlo.~\cite{pigcmc}
All of the simulations have been performed using $1.5 \times 10^6$ MC moves
per thermodynamic point, after $7.5 \times 10^5$ moves for equilibration.
We consider adsorption in an isolated nanotube of height $Z = 400$~\AA. Out of
100 Monte Carlo moves, 80 were insertion/deletion attempts and 20 were hybrid
MC moves.~\cite{hmc,pigcmc}

A first test on the efficiency of the Boltzmann bias method is performed by
calculating the adsorption of hydrogen in the (2,8) tube at $T=77$~K. The
results of the Boltzmann bias method are compared with the standard
prescription for insertion moves in Fig.~\ref{fig:numbers}(a), where it is
shown that the two methods lead to the same results as far as adsorption is
concerned.

The results for the same tube at the lowest temperature, shown in
Fig.~\ref{fig:numbers}(b) show a slight disagreement, especially in the region
around saturation.  The isotherms obtained with the two methods for the (3,6)
tube at $T=20$~K, reported in Fig.~\ref{fig:numbers}(c), are significantly
different. The origin of the difference is due to the poor performance of the
standard insertion method, that results in a very high rejection rate and
hence in a poor convergence towards the equilibrium density. The result of the
standard PIGCMC simulation of H${}_2$ adsorption in the (3,6) tube using twice
as many steps for equilibration and production, also reported in
Fig.~\ref{fig:numbers}(c), confirms this observation.  As is apparent
this second isotherm tends to converge to the correct result, but still shows
evident signs of poor equilibration.

\begin{figure}
\epsfig{file=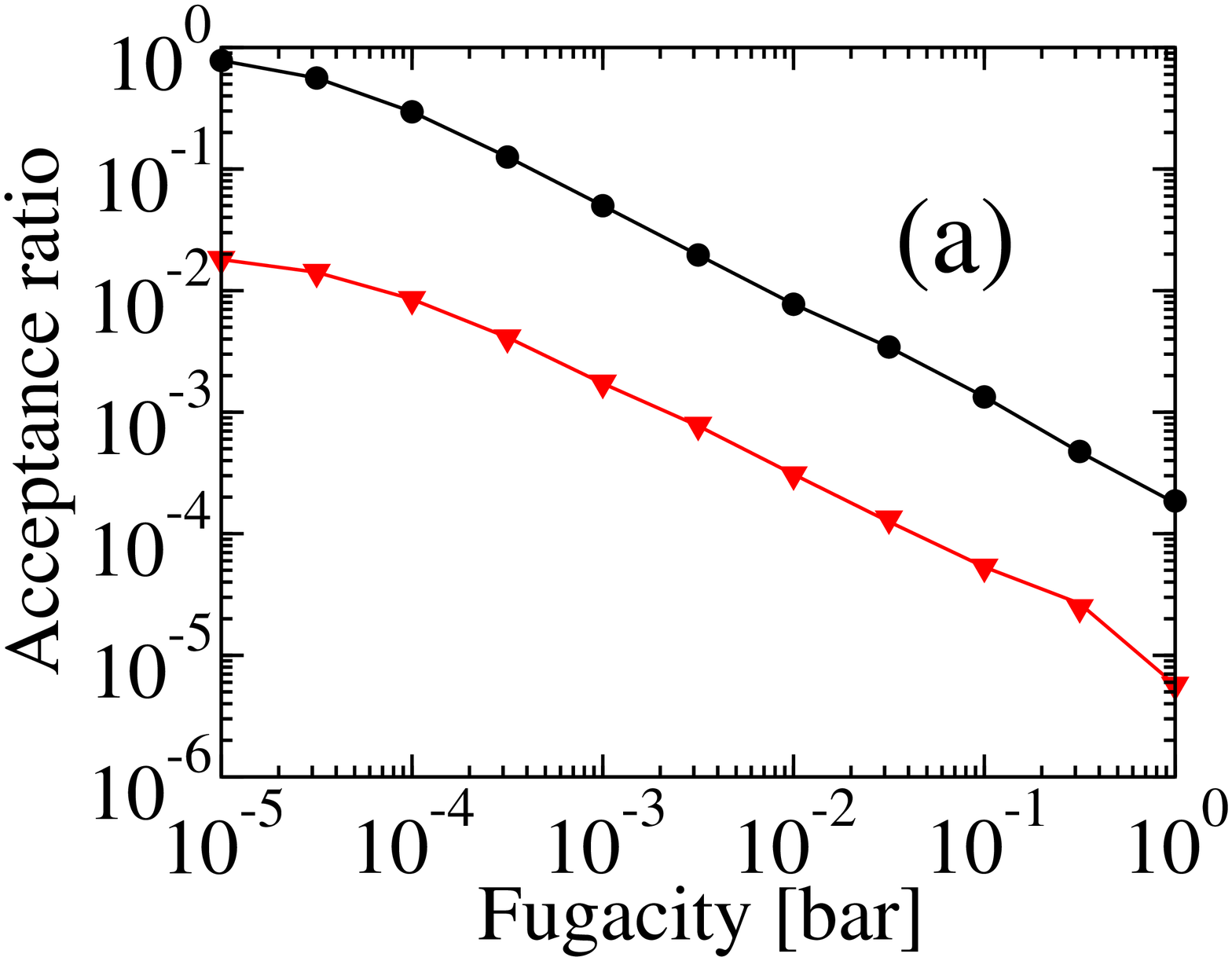,width=0.8\linewidth}
\epsfig{file=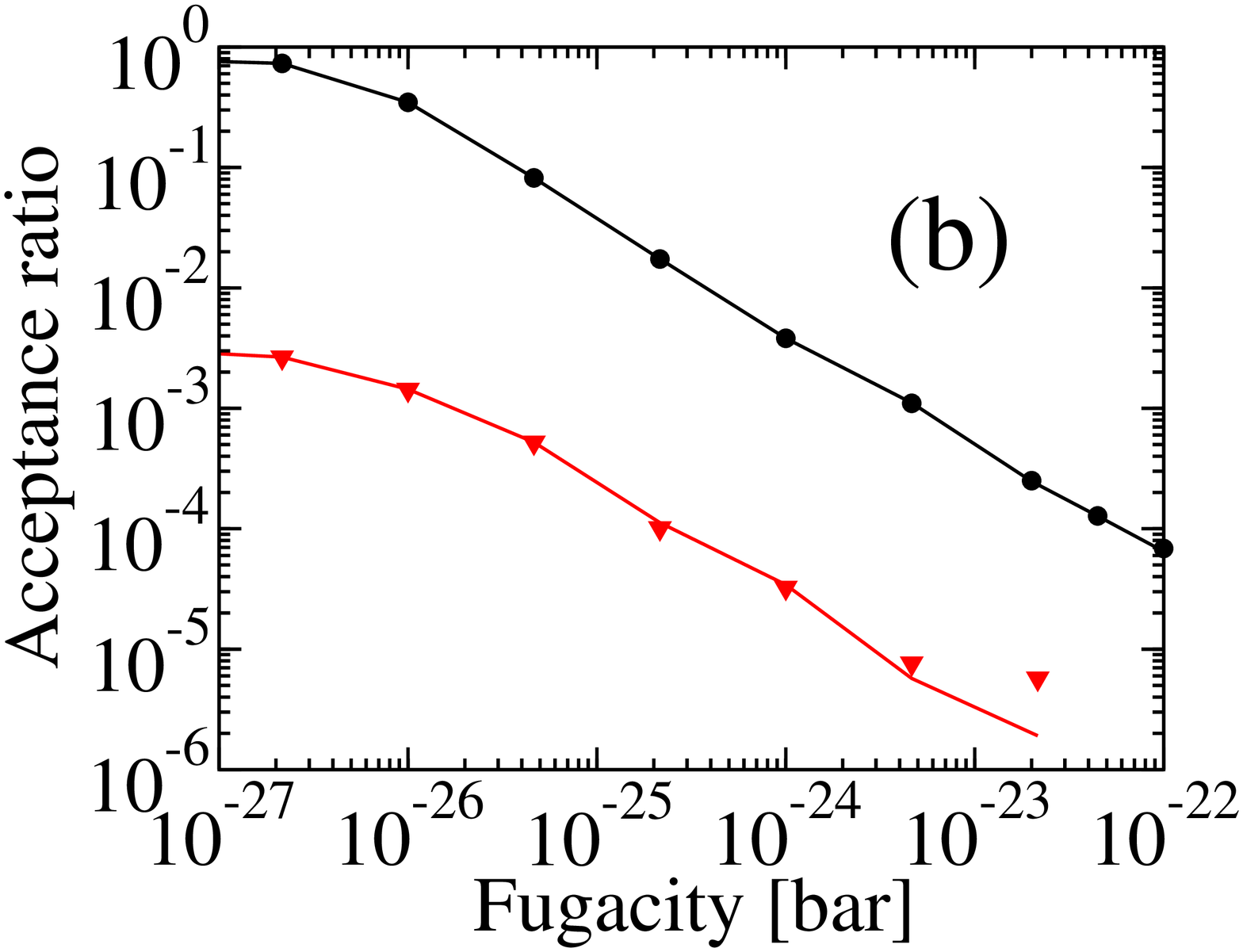,width=0.8\linewidth}
\epsfig{file=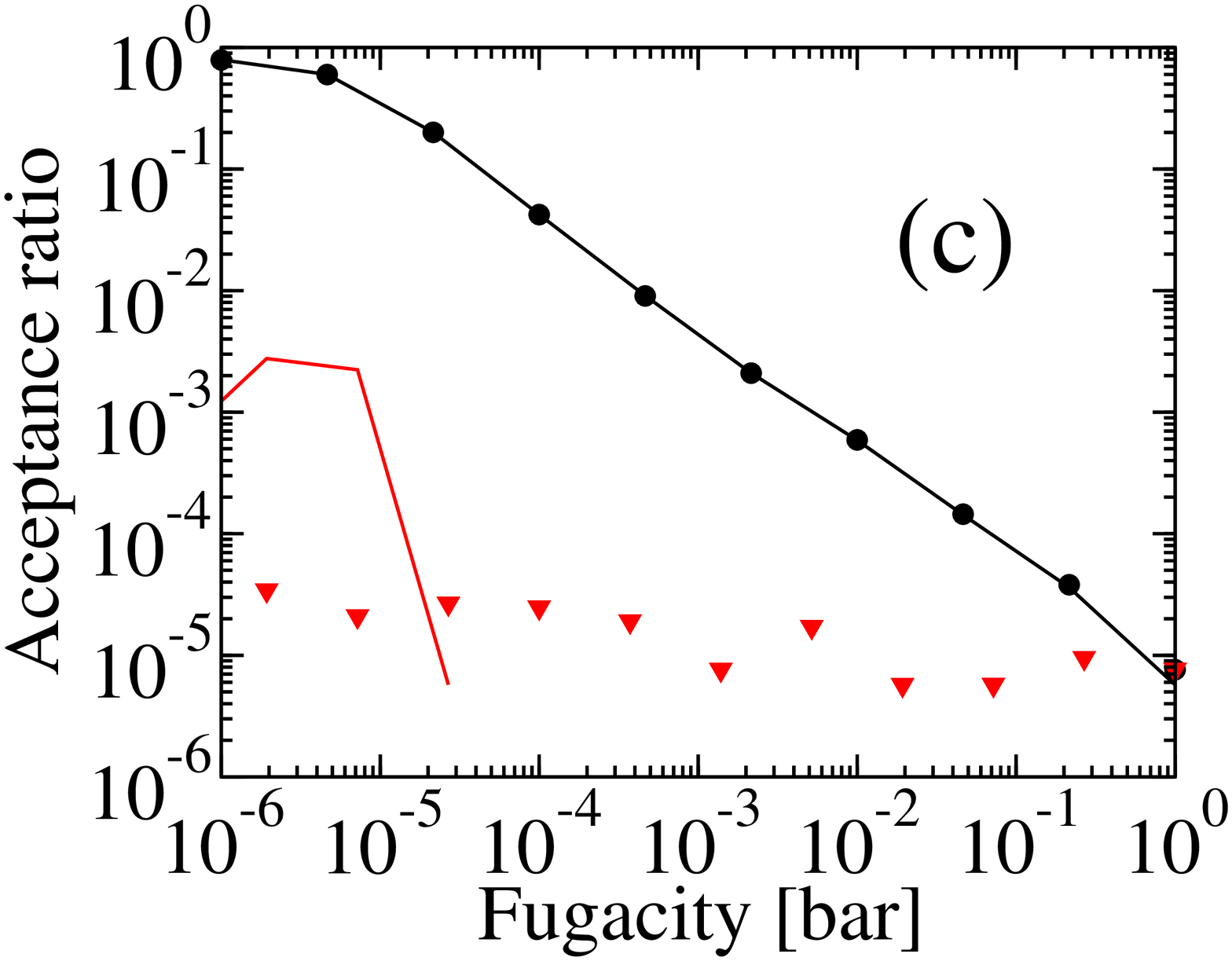,width=0.8\linewidth}
\caption{Insertion and deletion acceptance ratios for the production runs of
  the grand canonical simulations. (a) (2,8) tube at $T=77$~K, (b) (2,8) tube
  at $T=20$~K and (c) (3,6) tube at $T=20$~K. The circles denote the
  acceptance ratios obtained using the Boltzmann bias method, whereas the
  triangles are the acceptance ratios from simulations using ideal gas ring
  polymers as insertion candidates. The solid lines join the points
  corresponding to the deletion acceptance ratios (not highlighted by any
  symbol).}
\label{fig:insdel}
\end{figure}

In order to prove this fact, we show in Fig.~\ref{fig:insdel} the insertion
and deletion acceptance ratios for all the simulations we have performed.  

We notice that the insertion and deletion acceptance ratios have the same
values for each state point, except than in the case of the (3,6) tube
simulated using the standard practice of inserting ideal gas ring polymers. A
simulation run that achieves the same acceptance ratios for insertion and
deletion is well equilibrated. The fact that this situation is not verified in
the (3,6) tube using standard moves means that equilibration has not been
achieved even after $7.5 \times 10^5$ steps.

Under the same conditions the Boltzmann bias method does not show any sign
that convergence has not been reached, and we have checked that one obtains
the same isotherm using a larger number of Monte Carlo moves ($2.5 \times
10^6$ for production and half as many for equilibration, result not shown).

We notice that at each thermodynamic point that has been investigated the
acceptance ratios for the Boltzmann bias method are at least one to two order
of magnitude higher than the ones obtained using the standard method, and they
can be over four orders of magnitude higher than the standard method in very
confining geometries at low pressure, such as it happens in the case of
hydrogen adsorption at $T = 20$~K in (3,6) tube (see
Fig.~\ref{fig:insdel}(c)).

As might have been expected, the acceptance ratio decreases as the density
of adsorbed molecules is increased. This is due to the higher probability of
overlap between the particle to be inserted and the particles already present
in the simulation cell as the density of adsorbed molecules is increased. In
order to raise the acceptance ratio in this case, other kinds of biases can be
considered, such as the configuration-bias Monte Carlo.  In any case the
acceptance ratios for the Boltzmann bias method are one to two orders of
magnitude higher than the acceptance ratios obtained using ideal gas ring
polymers as candidates. This means that simulations using the Boltzmann bias
method need one to two order of magnitude less steps to achieve equilibration
as the standard method.

\subsection{Zero and finite pressure selectivity in narrow tubes}

\begin{table}
  \begin{tabular}{c|c|c|c}
Nanotube type & Mixture & Temperature (K) & Zero pressure selectivity \\
\hline
(3,6) & T${}_2$/H${}_2$ & 20 &  181000 \\
(3,6) & T${}_2$/H${}_2$ & 77 &  10.5   \\
(3,6) & D${}_2$/H${}_2$ & 20 &  5000   \\
(3,6) & D${}_2$/H${}_2$ & 77 &  5.5 \\
\hline
(2,8) & T${}_2$/H${}_2$ & 20 &  32   \\
(2,8) & T${}_2$/H${}_2$ & 77 &  1.55 \\
(2,8) & D${}_2$/H${}_2$ & 20 &  12.5 \\
(2,8) & D${}_2$/H${}_2$ & 77 &  1.40 \\    
  \end{tabular}
\caption{Zero pressure selectivities of hydrogen mixtures in carbon nanotubes}
\label{tab:nt_sel}
\end{table}

As is well known quantum diffraction effects result in different adsorption
properties of various isotopes of the same specie, a phenomenon which is
particularly evident at low temperatures and under conditions of strong
confinement.~\cite{KatorskiW64,Moiseyev75,krylov,wang,challa_zerop,challa,Garberoglio2006,goldfield2006,Garberoglio2007}

The differential adsorption of two species $A$ and $B$ is measured by the
selectivity $S(A/B)$, defined as  
\begin{equation}
S(A/B) = \frac{x_A/x_B}{y_A/y_B}
\label{eq:S}
\end{equation}
where $x_i$ is the molar fraction of specie $i$ in the adsorbed phase, and
$y_i$ is the molar fraction of the same specie in the gas phase, under
conditions of thermodynamic equilibrium.
When the pressure is so low that the fluid-fluid interaction between the
particles can be neglected, the selectivity can be written as a function of
the single particle energy levels of species $A$ and $B$ in the adsorbed
phase~\cite{challa}
\begin{equation}
S_0(A/B) = \left( \frac{m_B}{m_A} \right)^{3/2}
\frac{Q_A}{Q_B}
\label{eq:S0}
\end{equation}
where $m_i$ are the masses of the two species and $Q_i$ is the single particle
partition function
\begin{equation}
Q_i = \sum_k \exp\left( -\beta E^{(i)}_k \right)
\end{equation}
It can be seen that, in the limit $T \rightarrow 0$, the selectivity is a
function of the difference of the zero point energies of the two adsorbed
species.  Values of the zero pressure selectivity for mixtures of hydrogen
isotopes adsorbed in carbon nanotubes are reported in Tab.~\ref{tab:nt_sel}.
As might be expected, the selectivity is higher when the mass ratio between
the isotopes is higher (all other things being equal), and has a strong
temperature dependence. The values reported there can be compared with the
values reported in literature for molecular sieves (like zeolites) used in gas
separation, where typical values are in the range 10-100, depending on the
components of the mixture to be separated.~\cite{sel1,sel2}

Challa {\em et al.} have investigated the effect of finite pressure on the
selectivity, by performing multicomponent grand canonical path integral Monte
Carlo simulations and calculating the selectivity directly from the definition
of Eq.~(\ref{eq:S}). They noted a progressive inefficiency in the acceptance
ratios for insertion especially at high pressures, and hence were unable to
investigate the finite pressure selectivity in conditions of strong
confinement.

\begin{figure}
\epsfig{file=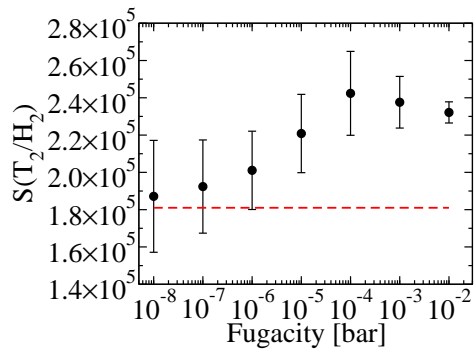,width=0.8\linewidth}
\caption{Selectivity for a T${}_2$/H${}_2$ mixture in the (3,6) carbon
  nanotube at $T=20$~K as a function of pressure. The simulation has been
  performed by assuming a mole fraction of T${}_2$ in the bulk phase equal to
  $y_{\rm{T}_2} = 5 \times 10^{-6}$. The horizontal dashed line corresponds to
the zero pressure value, reported in Tab.~\ref{tab:nt_sel}.}
\label{fig:S_3.6}
\end{figure}

The use of the Boltzmann biased algorithm allows one to gain a factor of 100
in the insertion efficiency, and correspondingly reduce the time needed for
equilibration. The results of our simulations for the pressure dependence of
the T${}_2$/H${}_2$ selectivity in the (3,6) nanotube are reported in
Fig.~\ref{fig:S_3.6}. We have used nanotubes of different length for different
pressures: the nanotube lengths were adjusted to have between 150 and 300
particles in the system after equilibration. In order to have an almost equal
amount of hydrogen and tritium within the simulation cell we have chosen to
perform the simulation by imposing a bulk mole fraction of T${}_2$ equal to
$y_{\rm{T}_2} = 5 \times 10^{-6}$.  We have used $2 \times 10^5$ steps for
equilibration and $4 \times 10^5$ steps for production at each state
point. The probability of performing an insertion/deletion has been set to
0.8.

One can see that, according to what has been observed in analogous confining
geometries, the selectivity actually increases from its zero pressure value 
and tends to saturate to a value which is some 40\% higher that the zero
pressure one.
This behavior can be understood using a simple mean field model by assuming
that, upon saturation, the particles are uniformly arranged within the tube,
so that each particle is confined also in the direction of the tube
axis. Using the simple theory of Eq.~(\ref{eq:S0}), the selectivity upon
saturation can be approximately written as the zero pressure selectivity times
the selectivity of a one dimensional harmonic oscillator corresponding to the
motion along the tube axis. A simple calculation shows that one does indeed
get the right magnitude of the increase.~\cite{Garberoglio2007}

\section{Adsorption and selectivity in slit pores} 

We have also applied our algorithm to the calculation of adsorption isotherms
and quantum sieving effects of hydrogen in a narrow slit pore. We have modeled
a slit pore as being formed by two graphite planes separated by a distance $H$
along the direction $z$, using the same parameters for carbon--hydrogen
interaction as in the case of nanotubes described above. We have obtained a
smooth solid-fluid interaction potential $v_{\rm slit}(z)$ by averaging the
values of the potential obtained by atomistic calculation over planes at
constant height $z$.  As a result of this approximation the quantum states of
a particle adsorbed within the slit pore are the product of free particle
states corresponding to the motion along the $xy$ plane and states coming from
the confined motion along the $z$ axis, which have been calculated by
numerical diagonalization of the 1D Hamiltonian in the free particle basis. We
have reached convergence by using the first 1024 plane waves.  We have chosen
to investigate in detail only the pore width of $H=5.7$~\AA, because this is
the narrowest slit still presenting a bound state for adsorbed hydrogen,
according to our model.

\begin{table*}
\begin{tabular}{c|c|c|c|c}    
Adsorbate & Temperature (K) & $\bar\mu /k_B$ (K) & 
Kinetic Energy/$k_B$ (K) & Potential Energy/$k_B$ (K)  \\
\hline
H${}_2$ & 20 & $-173.2$ & 205.4 & $-361.8$ \\
D${}_2$ & 20 & $-273.5$ & 147.7 & $-411.3$ \\
T${}_2$ & 20 & $-316.1$ & 123.0 & $-433.2$ \\
\hline
H${}_2$ & 77 & $-112.6$ & 262.4 & $-361.7$ \\
D${}_2$ & 77 & $-193.2$ & 205.1 & $-410.9$\\
T${}_2$ & 77 & $-222.4$ & 181.1 & $-432.2$ \\
\end{tabular}
\caption{Single particle properties of hydrogen isotopes confined in a slit
  pore with $H = 5.7$~\AA, obtained by direct diagonalization of the single
  particle Hamiltonian. The quantity $\bar{\mu}$ is defined in
  Eq.~(\ref{eq:mubar}).}
\label{tab:h2_slit}
\end{table*}

We report in Tab.~\ref{tab:h2_slit} the average kinetic and potential energies
of single hydrogen isotopes confined within the $H=5.7$~\AA\ slit pore. We
notice that the average potential energies do not change very much with
increasing temperature, a signature of the fact the the first excited state is
very well separated from the ground state for all the three isotopes. The
smallest separation is observed for T${}_2$ and is of the order of $\Delta E /
k_B = 412$~K. As a consequence the hydrogen isotopes have a very low probability of
being in the first excited state at the temperatures used in this study, and
the temperature dependence of the kinetic energies reported in
Tab.~\ref{tab:h2_slit} comes from the different average kinetic energies
corresponding to the motion along the $xy$ plane.

\begin{figure}
\epsfig{file=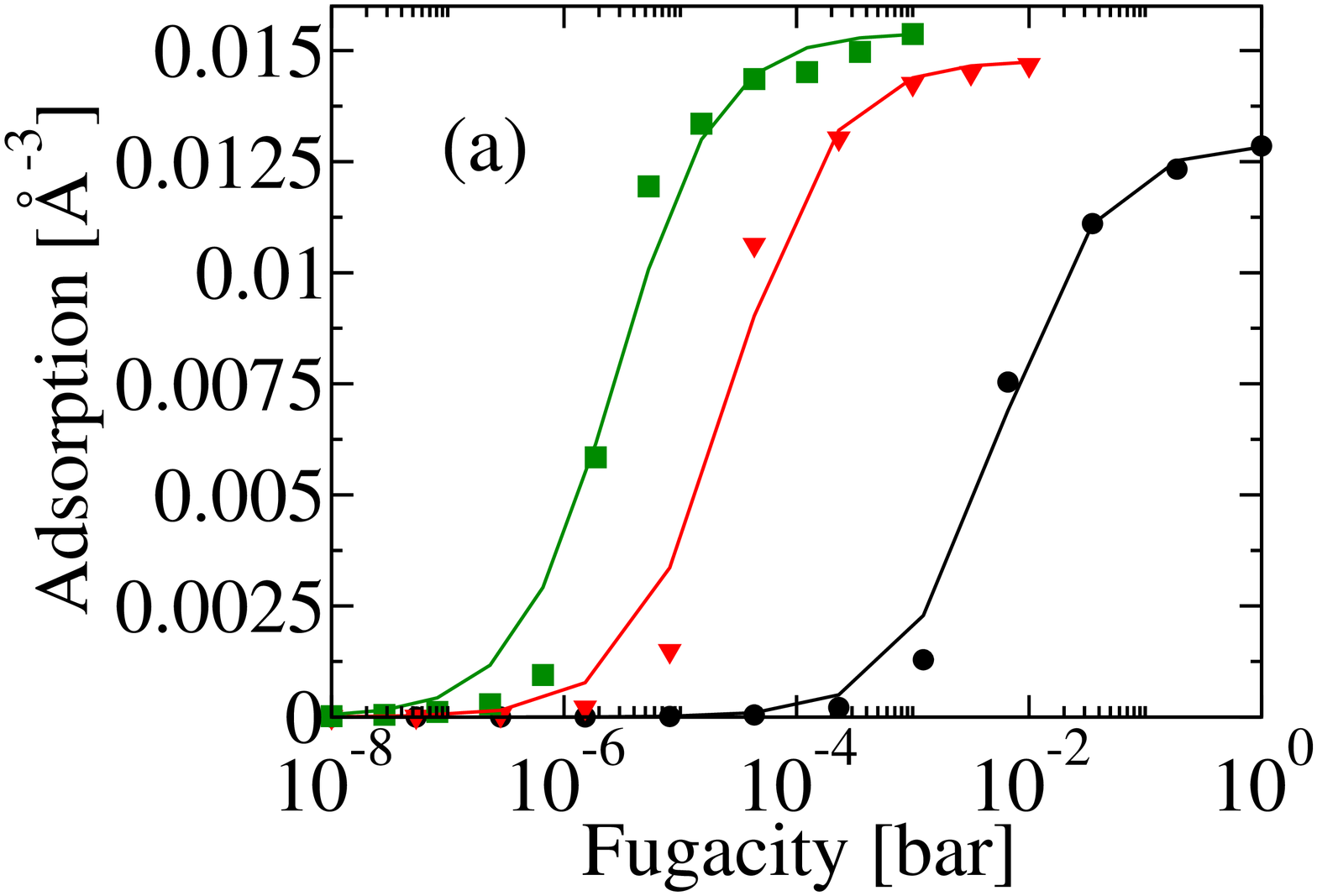,width=0.8\linewidth}
\epsfig{file=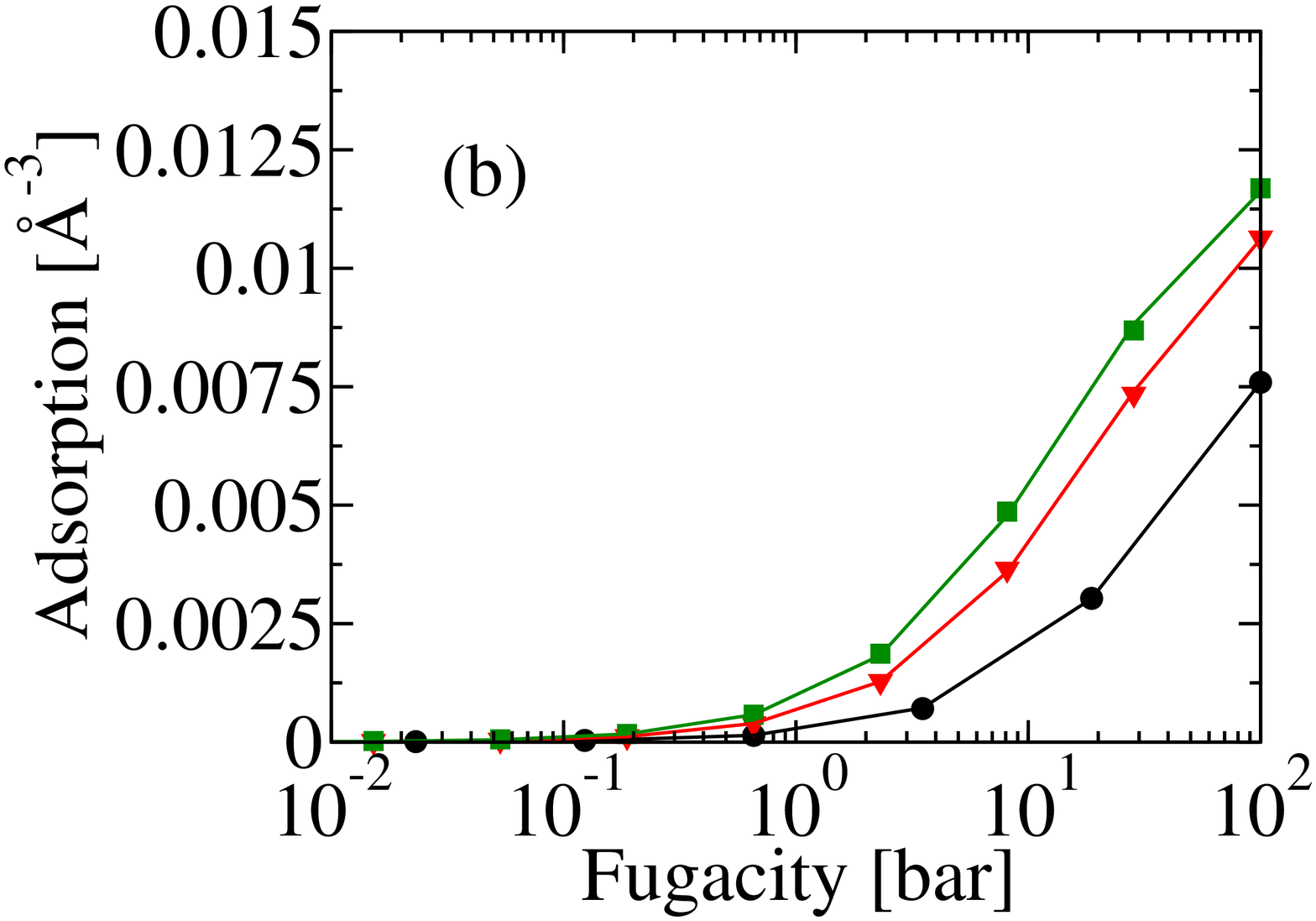,width=0.8\linewidth}
\caption{Isotherms of adsorption of hydrogen isotopes within a slit pore with
  $H=5.7$~\AA\ at $T=20$~K (a) and $T=77$~K (b). Circles: H${}_2$, triangles:
  D${}_2$, squares: T${}_2$. The adsorbed density is obtained by dividing the
  total number of molecules in the simulation cell by the product of the
  graphite planes' surface and the distance $H$ between them. The lines are a
  fit with a Langmuir type isotherm.}
\label{fig:slit_iso}
\end{figure}

The difference between the potential and kinetic energies of the various
isotopes confined within the slit pore already indicates that a strong
isotope effect is to be expected upon adsorption.  In fact the the adsorption
isotherms reported in Fig.~\ref{fig:slit_iso} show that this is indeed the
case. 

The isotherms have been calculated using the PIGCMC technique and using, for
all the isotopes, the same number of beads $P$ as in the simulation of
adsorption within carbon nanotubes. For each pressure we have performed $2.5
\times 10^5$ PIGCMC steps for equilibration, followed by $5 \times 10^5$ steps
for production. The lateral dimensions of the pore were such that in
saturation condition, at least 300 particles were present in the simulation
box, resulting in a size of about $L = 70$~\AA.

The computed isotherms for the three isotopes at $T=20$~K have a similar
shape, though they are separated by two or three orders of magnitude in
pressure; the heaviest specie is adsorbed at lower pressures than the lightest
one.  Moreover the saturation density is markedly different for the three
isotopes, with T${}_2$ having the largest and H${}_2$ the smallest.  The
isotope effect is also sizeable for the adsorption at $T=77$~K, though not as
significant as it is at the lowest temperature.  The adsorption isotherms
plotted in Fig.~\ref{fig:slit_iso} are in good quantitative agreement with
analogous computer simulation results already appeared in the
literature.~\cite{slit-karl,slit-fh}

\begin{table}
\begin{tabular}{c|c|c}
Mixture & Temperature (K) & Zero pressure selectivity \\
\hline
T${}_2$/H${}_2$ & 20 & 1263 \\
T${}_2$/H${}_2$ & 77 & 4.3  \\
D${}_2$/H${}_2$ & 20 & 151  \\
D${}_2$/H${}_2$ & 77 & 2.8  \\
\end{tabular}
\caption{Zero pressure selectivities of hydrogen mixtures in the
  $H=5.7$~\AA\ slit pore}
\label{tab:slit_sel}
\end{table}

Calculation of the selectivity confirms the presence of a significant isotope
effect. The zero pressure selectivities, shown in Table~\ref{tab:slit_sel}, are
found to be quite high at low temperature, reaching a value of around $1200$
for the case of T${}_2$/H${}_2$ mixtures and being of the order of $150$ for
D${}_2$/H${}_2$ mixtures.  The selectivity has a very strong dependence on the
temperature, and drops to values around $4$ and $3$ for T${}_2$/H${}_2$ and
D${}_2$/H${}_2$ mixtures, respectively, at a temperature of
$T=77$~K.~\cite{slit-pores}

It is interesting to notice that the values of the zero pressure selectivity
calculated for the narrow slit pores, although lower than the ones of the
narrowest carbon nanotubes, are nonetheless larger than those one could expect
on the basis of geometrical considerations, given the 2D confining nature of
the carbon nanotubes and the 1D confining nature of slit pores.  In fact, if
one assumes that the two confining directions are independent the single
particle partition function can be written as the product of the partition
functions corresponding to the two degrees of freedom. As a consequence of 
Eq.~(\ref{eq:S0}), then, the selectivity could be written as the product of
two contributions, one for each confining direction.
Therefore the selectivity of the nanotube should be equal to the square of the
selectivity of the slit pore, if the two systems have the same confinement
property along each confining direction. 

\begin{figure}
\epsfig{file=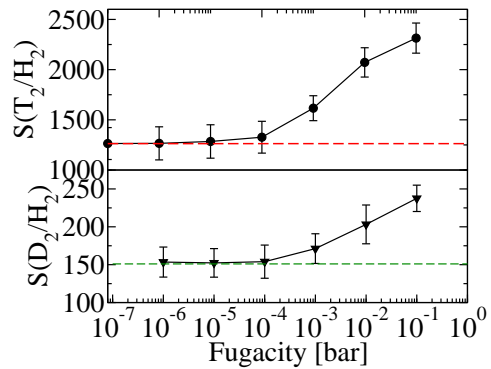,width=0.8\linewidth}
\caption{Selectivity for hydrogen isotope mixtures as a function of pressure
  in the $H=5.7$~\AA\ slit pore at $T=20$~K. The upper panel shows the
  selectivity of a T${}_2$/H${}_2$ mixture ($y_{\rm T_2} = 8 \times 10^{-4}$),
  and the lower panel the selectivity of a D${}_2$/H${}_2$ mixture ($y_{\rm
  D_2} = 7 \times 10^{-3}$). The horizontal dashed lines correspond to the
  zero pressure values, reported in Tab.~\ref{tab:slit_sel}.}
\label{fig:slit_sel_20K}
\end{figure}

\begin{figure}
\epsfig{file=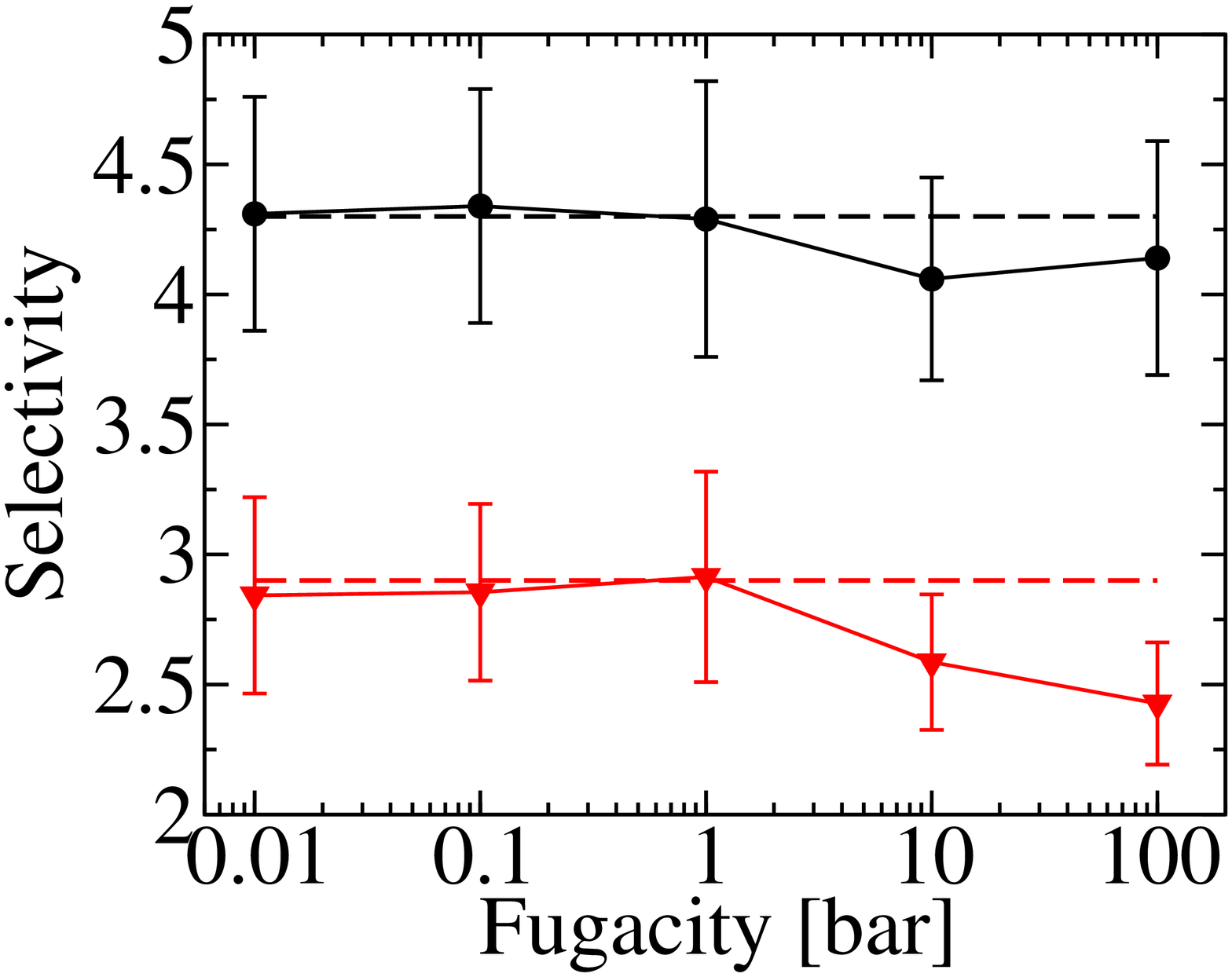,width=0.8\linewidth}
\caption{Selectivity for hydrogen isotope mixtures as a function of pressure
  in the $H=5.7$~\AA\ slit pore at $T=77$~K. The circles shows the 
  selectivity of a T${}_2$/H${}_2$ mixture ($y_{\rm T_2} = 0.23$),
  and triangles the selectivity of a T${}_2$/H${}_2$ mixture ($y_{\rm
  D_2} = 0.35$) as a function of the external pressure. The horizontal dashed
  lines correspond to the zero pressure values, reported in
  Tab.~\ref{tab:slit_sel}.}
\label{fig:slit_sel_77K}
\end{figure}

We have calculated the selectivity as a function of pressure for adsorption of
hydrogen isotopes in slit pores using the Boltzmann bias method.  We show in
Fig.~(\ref{fig:slit_sel_20K}) and (\ref{fig:slit_sel_77K}) the results
obtained at the temperature of $T=20$~K and $T=77$~K respectively.  The
calculations at each pressure have been performed by simulating the adsorption
of mixtures using the Boltzmann bias path integral grand canonical method and
calculating the finite pressure selectivity directly from its definition,
Eq.~(\ref{eq:S}).
For each state point a number between $1.5 \times 10^5$ and $5 \times 10^5$ of
Monte Carlo steps were performed for the equilibration phase (depending on the
pressure) and twice as many steps for the production run.  In order to have an
almost equal number of particles for both species within the simulation cell,
we set the bulk mole fractions for the isotopes to the values indicated in the
figures' caption.

The results for $T=20$~K are qualitatively similar to what has already been
observed for strongly confining nanotubes:~\cite{challa,Garberoglio2007} the
selectivity rises from its zero pressure value for both tritium/hydrogen and
deuterium/hydrogen mixtures, almost doubling its zero pressure value.  This is
consistent with the mechanism outlined above in the case of carbon
nanotubes. Upon reaching saturation each adsorbed particle is confined also
in the $xy$ plane and one would expect the selectivity to raise by a factor
corresponding to the square of what is observed in the case of nanotubes.
Since the selectivity increases by a factor of 1.4 in the case of nanotubes,
one expects the selectivity in the slit pores to increase by a factor of $1.4
\times 1.4 \simeq 2$, as is indeed observed from the simulation result at $T =
20$~K.

At the higher temperature of $T=77$~K the pressure dependence of the
selectivity is less dramatic. One can see from Fig.~\ref{fig:slit_sel_77K}
that the selectivity remains almost constant over four decades in pressure,
possibly showing a slight tendency to decrease from its zero pressure value,
as is particularly evident in the case of the deuterium/hydrogen mixture.
In this case the further confinement along the $xy$ plane is less efficient 
than at low temperature, and the selectivity remains almost constant.

\section{Conclusions} 

We have developed an efficient method for the insertion/deletion moves of ring
polymers in path integral grand canonical Monte Carlo simulations of fluids
adsorbed in strongly confining geometries.  In this case the kinetic energy of
the adsorbed phase is larger than the thermal kinetic energy of a perfect gas
at the same temperature and the standard procedure of inserting ring polymers
configuration drawn from the ideal gas distribution is highly inefficient
because the polymers are on average too large to fit within the narrow
channels, resulting in a very low acceptance ratios.  The new algorithm
presented here is based on the idea of taking ring polymer configurations
from a canonical simulation of independent particles already adsorbed.

The efficiency of the method comes from the fact that one uses as insertion
candidates ring polymers already equilibrated according to the confining
solid-fluid potential. We have observed an increase of the acceptance ratio
for insertion moves up to 2 order of magnitudes when compared with the
standard practice of trying to insert ideal gas ring polymer configurations,
and a corresponding decrease of the time needed to perform a simulation.

The algorithm has been validated by calculating adsorption isotherms of pure
fluids within narrow carbon nanotubes and graphite slit pores and comparing
them with results already published in the literature, or obtained by us using
the standard algorithm.  Moreover we have calculated the pressure dependence
of the isotopic selectivity in narrow nanotubes and slit pores and showed that
even in the case of very strong confinement its value increases from the zero
pressure limit.

Although this method becomes progressively inefficient near saturation
conditions, its simplicity, effectiveness and ease of programming makes it a
suitable candidate for simulating adsorption of quantum gases under conditions
of strong confinement in a wide range of loadings.

Finally we would like to point out that this simulation technique can be
applied to classical simulation of adsorption of small molecules or polymers
within narrow channels.~\cite{polypeptide-ads}

\begin{acknowledgments}
The author thanks Dr. F. Pederiva for useful discussions and
Prof. R. Vallauri for a careful reading of the final manuscript.

The computer simulations have been performed on the HPC facility {\sl Wiglaf}
at the Physics Department of the University of Trento.
\end{acknowledgments}

\appendix

\begin{widetext}
\section{Probability distribution for ring polymer configuration}
\label{sec:appendix}

The partition function of a free particle is
\begin{equation}
Z = \int d^3x_1 ~ \langle x_1 | \exp(-\beta p^2/ 2 m) | x_1 \rangle = 
\frac{V}{\Lambda^3}
\end{equation}
that becomes, after a Trotter expansion,
\begin{equation}
Z = \int d^3x_1 d^3x_2 \ldots d^3x_P ~ 
\langle x_1 | \exp(-\beta p^2/ 2 m P) | x_2 \rangle
\ldots
\langle x_P | \exp(-\beta p^2/ 2 m P) | x_1 \rangle  
\end{equation}  
The integral can be rewritten using the variables defined in
Eq.~(\ref{eq:change_var})
together with the definition $\Delta r_P = - \displaystyle\sum_{i=1}^{P-1}
\Delta r_i = x_1 - x_P$, 
\begin{equation}
Z = \left(\frac{P^{3/2}}{\Lambda^3} \right)^P 
\int d^3r_1 d^3\Delta r_1 \ldots d^3 \Delta r_{P-1}
\exp \left[ -\frac{K}{2} \left(
\Delta r_1^2 + \Delta r_2^2 + \ldots + \Delta r_{P-1}^2 + \Delta r_P^2
\right) \right]
= \frac{V}{\Lambda^3}  
\end{equation}  
where $K=2 \pi P/ \Lambda^2$. Since the integrand does not depend on $r_1$ on
can integrate it obtaining
\begin{equation}
1 = \Lambda^3 \left(\frac{P^{3/2}}{\Lambda^3} \right)^P 
\int d^3\Delta r_1 \ldots d^3 \Delta r_{P-1}
\exp \left[ -\frac{K}{2} \left(
\Delta r_1^2 + \Delta r_2^2 + \ldots + \Delta r_{P-1}^2 + \Delta r_P^2
\right) \right]  
\end{equation}  
which can be interpreted as the normalization condition on the probability of
observing a ring polymer of $P$ beads with separation from the neighboring
beads equal to $\Delta r_1 \ldots \Delta r_{P-1}$, the last one being fixed by
the condition of having a {\em closed} ring polymer.
Hence the probability for such a configuration is
\begin{equation}
F_{\mathrm{ring}} (\Delta r_1, \ldots, \Delta r_{P-1}) = 
\Lambda^3 \left(\frac{P^{3/2}}{\Lambda^3} \right)^P 
\exp \left[ -\frac{K}{2} \left(
\Delta r_1^2 + \Delta r_2^2 + \ldots + \Delta r_{P-1}^2 + \Delta r_P^2
\right) \right]
\label{eq:prob_ring}
\end{equation}  
\end{widetext}

\bibliography{bb-pimc}

\newpage

\end{document}